# COMMENTS REGARDING WILLIAM HERSCHEL APRIL 1787 REPORT OF AN ERUPTING VOLCANO ON THE MOON: WERE THESE OBSERVATIONS THE MANIFESTATION OF IMPACT MELT, PRODUCED BY A METEORITE FROM THE LYRID METEOR SHOWER?

William. Bruckman[1], and Abraham. Ruiz[1] ; [1]University of Puerto Rico At Humacao, Department of Physics and Electronic. (miguelwillia.bruckman@upr.edu) (abraham.ruiz@upr.edu),

**ABSTRACT:**
We consider that the report by William Herschel on April 19 and 20, 1787, about an erupting volcano on the Moon, were really the observations of impact melt, produced by a meteorite from the Lyrid meteor shower. According to our investigation the probable resulting crater of this event is a Lunar Cold Spot, with coordinates similar to those given in Herschel`s report, and we also argue that this impact has very young age characteristics.

In 1787 William Herschel wrote [1] "April 19, 10h. 36` sidereal time: I perceive three volcanoes in different places of the dark part of the new moon. Two of them are either already nearly extinct, or otherwise in a state of going to break out; which perhaps may be decided next lunation. The third shows an actual eruption of fire, or luminous matter. I measured the distance of the crater from the northern limb of the moon, and found it to be 3` 57".3. Its light is much brighter than the nucleus of the comet which M. Mechain discovered at Paris the 10[th] of this month. ; April 20, 10h. 2` sidereal time: The volcano burns with greater violence than last night. I believe its diameter cannot be less than 3``, by comparing it with that of the Georgian planet; as Jupiter was near at hand, I turn the telescope to his third satellite, and estimated the diameter of the burning part of the volcano to be equal to at least twice that of the satellite. Hence we may compute that the shining or burning matter must be above three miles in diameter. It is of an irregular round figure, and very sharply defined on the edges…..The appearance of what I have called the actual fire of eruption of a volcano, exactly resembled a small piece of burning charcoal, when it is covered by a very thin coat of white ashes, which frequently adhere to it when it has been some time ignited; and it had a degree of brightness, about as strong as that with which such a coal would be seen to glow in faint daylight….."

Nowadays we do not believe in active volcanoes on the moon, but then, what was William Herschel observing in such detail, and for several days? . An astronomer as important deserves the benefit of the doubt. Here we will consider that his observation can be explained by a meteorite that created impact melt in the area. Furthermore, if so, then, this event was probably produced by a large fragment from the Lyrid meteor shower, in which grains and chunks of comet Thatcher orbit hit the Earth and the Moon with velocities near 50km/s. This happens every year about between the 16[th] to the 26[th] of April, and the maximum intensity at the time of Herschel observations occurred on, or very close to April 20 ( see for example: https://www.bashewa.com/wxmeteor-showers.php?shower=Lyrids&year=1787; http://meteorshowersonline.com/lyrids.html ).

We would like to point out the relevant hypothesis that a fragment from the Encke comet caused the Tunguska event [2], because it is another example of a possible connection between a meteor shower and a large meteorite.

The conventional wisdom is that small meteorites impacts on the moon could be visible only for few seconds. These events are called lunar flashes, and are historically associated with the topic of Lunar Transient Phenomenon. Incidentally, an account of a "dazzling white star" observed in the disc of the moon on April 24,1874, seemingly describes one historical lunar flash, that may have been a Lyrid lunar meteorite: https://www.astrobio.net/retrospections/the-curious-history-of-the-lyrid-meteor-shower/. Nevertheless, an impact need not be short-lived, for if it generates melting then, this may produce sufficient visible light to be observable for longer periods. Since initial investigations by Shoemaker et al [3], Howard and Wilshire [4], and Hawke and Head [5] to the more recent research it is shown that impact melt can occur, in and near craters, from large sizes to the sub kilometer diameters situations ( see for example Stopar et al [6] ). These studies tell us that the thermal and mechanical dynamics after an impact are very important in understanding lunar surface characteristics. For instance, quoting reference [7]: " LROC images suggest that new flows can emerge from melt ponds an extended time period….Nevertheless, despite the almost instantaneous nature of impact melt generation and initial emplacement, we conclude that impact melts in and around craters are compound deposits created by multiple stages flow". Thus, if after an initial impact we have sufficiently high temperatures to attain red visible light, then perhaps Herschel`s burning charcoal description would be an intriguing and plausible account of an impact melt event.

According to Herschel, the erupting object was happening a distance of 3` 57".3 of the northern limb of the Moon. Interpreting the northern limb of the

Moon as meaning near the north pole, then we find an interesting candidate, which is a lunar cold spot impact crater, about 4` from the pole, at latitude 45.985 and longitude -14.715 degrees (diameter, D, about 750 meters, see Figures 1,2,3,4,5 ). Cold spots (Bandfield et al [8]) are a family of very young lunar craters, that are characterized by having a surface, surrounding the impact, with nighttime average temperature at least 2k less than the background, rock free, regolith temperature. This property seems common to all new and small impacts ( 100m $\approx \leq D \leq \approx$1.5km ), but ephemeral, so that their total number is only about 4,000 [8]. However, the area interior to the above candidate crater and its immediate environment, up to few diameters, have larger than average temperatures, which is consistent with the presence of boulders in this area (Figure 2). This temperature duality happens when the cold spot craters are large enough. Boulders and warmer interior areas could also be associated with melt events [9].

If this candidate crater was what Herschel observed in 1787, we should look for signs of a pristine surface near it, and this can be done by inspecting his ejecta, while expecting it to be clean of new impacts. For the relatively small crater considered here, with not enough displaced material, we have that a newly formed crater would not be easy to differentiate from the already existing old ones (see Figure 6, left). In this situation we can use a procedure [10], based on the recognition of the effect of a new impact on the ejecta blanket around a fresh crater, visible on small or medium angle solar incidence photos of the Lunar Reconnaissance Orbiter Camera (LROC). In low solar incidence LROC photos the new impacts usually appear as scars, over the background (Figure 6, right) with symmetrical shapes that break the original impact radial symmetry; thus these patterns distinguish them from old impacts. Hence, by observing the corresponding (LROC) images, at a large angle of solar incidence, covering the same location, we should be able to identify the new impact crater that is responsible for the peculiar distinguishable and magnified scar event. The above method was used to establish the relatively young age of a cold spot in Lunar Mare Fecunditatis (latitude 3.635 and longitude 48.929 degrees) , a result consistent with the fact that this crater is not found in 1938 photos (E3 and D3) of the Kuiper et al, The Consolidated Lunar Atlas [11] (see Figures 7a and 8a ), although it is quite clear in LROC photos (Figures 7b and 8b). The above procedure is more effective with small solar incidence angle photos, but for the Herschel candidate crater under consideration here we found only LROC medium solar incidence angle photos. Nonetheless, no clear evidence of any new impact was found, thus suggesting that it is indeed a young crater relative to other cold spots. It will be valuable to extend this study using very low solar incidence LROC photos.

Interestingly, we also note (see Figure 5) signs of impact melt. For instance, the arrow in Figure 5 points to what apparently is a ray of material spraying from the interior of the crater to over the rim, that, perhaps, can be interpreted as original melted material ejected. However, it is for expert eyes to verify such probable interpretation.

We know, from studies of impacts on Earth and the Moon, that a near megaton energy impact is capable of forming a several hundred meter diameter crater on the moon, similar to the one in Figure 4. Furthermore, the rate of these impacts on our planet is estimated to be approximately one every 15 years ( Silber et al [12], Bruckman et al [13] ). That rate translates for the lunar surface as about one impact every couple of centuries. Therefore, our interpretation of Herschel`s description as a possible meteorite does not disagree with the above impact rate estimates.

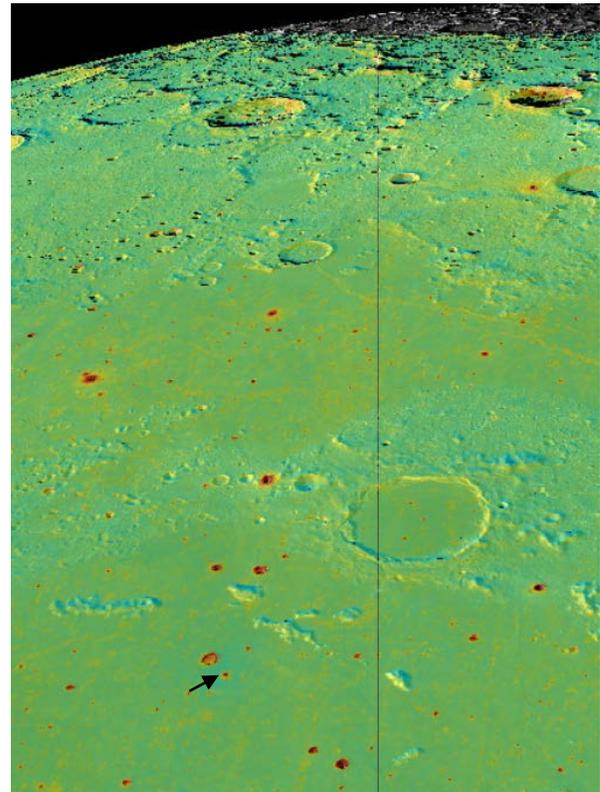

Figure 1: The cold spot (arrow) at latitude 45.985 and longitude -14.715. LROC Diviner view. Blue colors denote lower temperatures, and red colors higher.

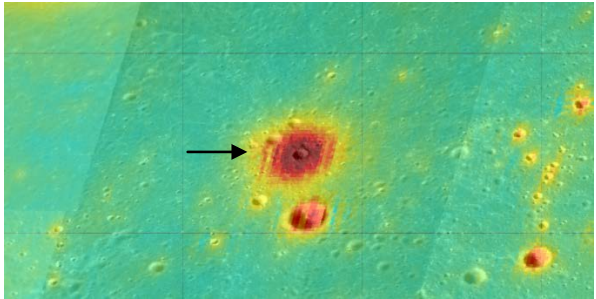

Figure 2: Amplification of the cold spot of Figure 1.

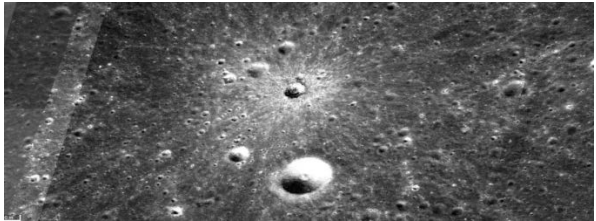

Figure 3: The cold spot, in Figure 2, view with LROC medium solar incidence.

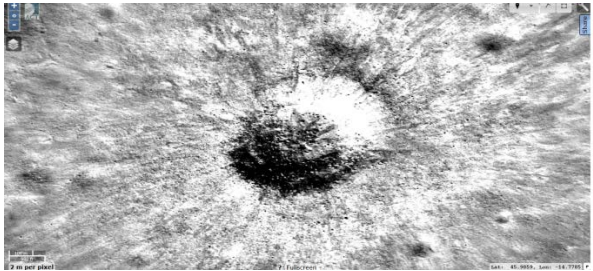

Figure 4: Amplification of Figure 3.

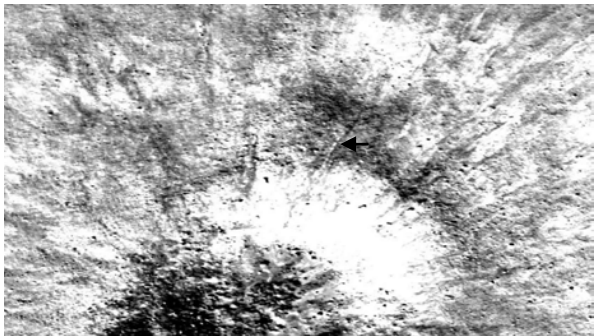

Figure 5: Amplification of Figure 4. The arrow shows an Intriguing ray.

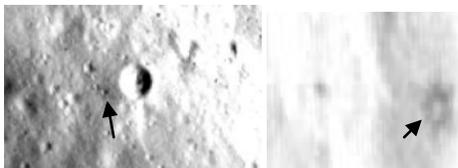

Figure 6: Arrow at left points to a new formed crater, latitude 2.452, longitude 45.437 viewed with large solar incidence; arrow at right points to the same crater, viewed with low solar incidence. Source LROC

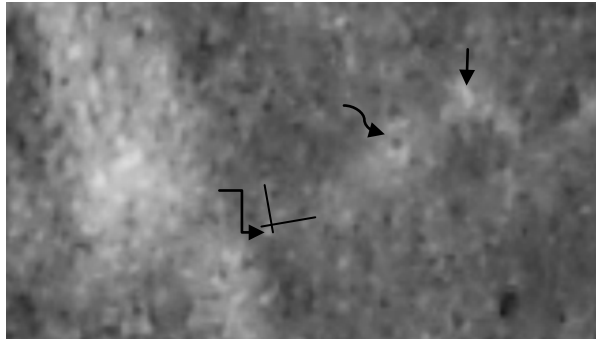

Figure 7a: The three arrows are pointing to objects in the 1938 E3 Consolidated Lunar Atlas Image, that are identified in LROC Figure 7b.

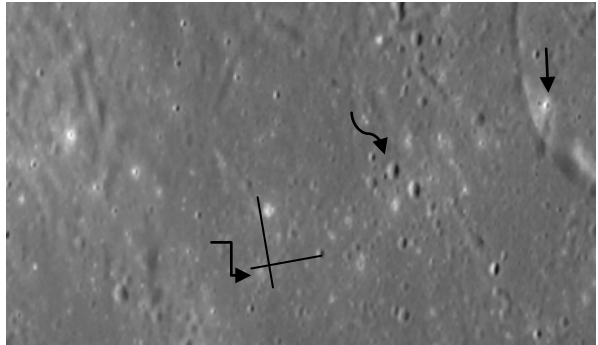

Figure 7b: LROC image of the region of Figure 7a.

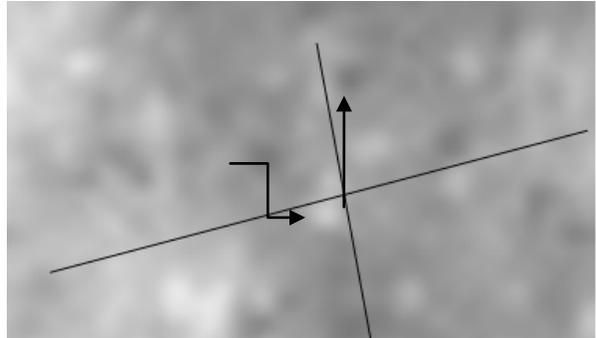

Figure 8a: Close-up of Figure 7a.

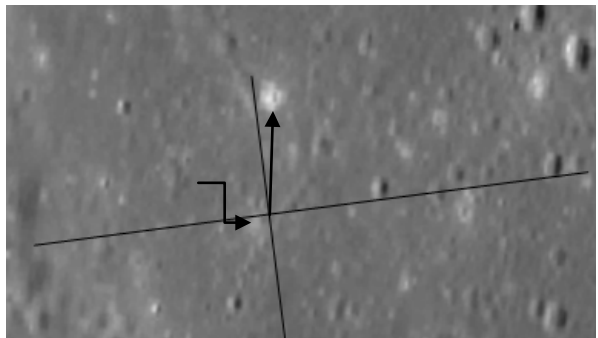

Figure 8b: Close-up of Figure 7b